\documentclass[intlimits,twoside,a4paper]{article}

\usepackage{graphicx}
\usepackage{amsmath,amssymb}
\usepackage[T2A]{fontenc}
\usepackage[cp1251]{inputenc}
\usepackage[eqsecnum]{cmpj2}

\issue{2015}{18}{3}{33702}
\doinumber{10.5488/CMP.18.33702}

\title[Quasiparticle electronic band structure of the alkali metal chalcogenides]%
{Quasiparticle electronic band structure of the alkali metal chalcogenides}
\author[S.V.~Syrotyuk, V.M.~Shved]{S.V.~Syrotyuk, V.M.~Shved}
\address{Lviv Polytechnic National University, 12 S. Bandera Str., 79013 Lviv, Ukraine}

\date{Received December 29, 2014, in final form March 16, 2015}
\begin{document}

\maketitle

\begin{abstract}

The electronic energy band spectra of the alkali metal chalcogenides M$_2$A (M: Li, Na, K, Rb; A: O, S, Se, Te) have been evaluated within the projector augmented waves (PAW) approach by means of the ABINIT code. The Kohn-Sham single-particle states have been found in the GGA (the generalized gradient approximation) framework. Further, on the basis of these results the quasiparticle energies of electrons as well as the dielectric constants were obtained in the GW approximation. The calculations  based on the Green's function have been originally done for all the considered M$_2$A crystals, except Li$_2$O.

\pacs 71.15.Mb, 71.15.Ap, 71.15.Nc, 71.20.Nr, 78.20.Ci, 71.15.Qe
\end{abstract}

\section{Introduction}

The alkali metal chalcogenides M$_2$A (M: Li, Na, K, Rb; A: O, S, Se, Te) are found to crystallize in the cubic anti-fluorite (anti-CaF$_2$-type) structure at ambient conditions. They draw considerable attention of researchers due to their possible applications in power sources, fuel cells, gas-detectors and ultraviolet space technology devices~\cite{b1}.

The properties of the crystals M$_2$O have been extensively studied experimentally~\cite{b2}, whereas the sulfides, selenides and tellurides of alkali metal have received less experimental attention. The electronic energy band spectra of the M$_2$A crystals  have been evaluated using the full potential linearized augmented plane waves plus local orbitals (FP APW$+$lo) method based on DFT~\cite{b1}. However, it is well known that the resulting band gap values in this approach are much underestimated. A proper way of calculating single-particle excitation energies or quasiparticle energies is provided by the Green's function theory. Here, the GW approximation (GWA) is used, which is the simplest working approximation beyond the Hartree-Fock approach taking screening into account~\cite{b3}.

Calculations of the electron energy spectrum beyond the local (LDA) or quasilocal (GGA) approximations were made only for the crystal Li$_2$O ~\cite{b4}. The Kohn-Sham ground-state data have been evaluated~\cite{b4} on the norm-conserving pseudopotential basis. On this basis there were obtained quasiparticle corrections to the eigenenergies using the GWA.

Then, the Bethe-Salpeter equation, which includes the screened electron-hole interaction as well as the unscreened electron-hole exchange term~\cite{b4}, was solved and the lowest exciton eigenvalue at 6.6~eV was found. This value is well compared with the optical absorption energy at about 6.6~eV. The GW corrections open the gap at the $\Gamma$  point by 2.1~eV yielding a minimum direct gap of 7.4~eV. Therefore, the difference between the GW energy and the excitonic energy gives the exciton binding energy of 0.8~eV.

The above listed applications of the alkali metal chalcogenides do not exhaust the potential capabilities of these crystals. In fact, the recently registered patents suggest a possible use of these crystals, doped with $d$- or $f$-transition elements, in spintronics~\cite{b5}. Finally, it is worth to mention an interesting theoretical prediction of the occurrence of a ferromagnetic half-metallic ordering in these crystals caused by the doping with nonmagnetic elements C, Si, Ge, Sn and Pb~\cite{b6}.

The compounds considered here have large lattice constants. As a result, the hybridization between respective orbitals of an impurity and a parent atom would be weak. Thus, the alkali metal atom, such as K, Na, Li or Rb can be substituted with each of the 3$d$, 4$d$ and 5$d$ transition metal elements and the rare-earth 4$f$ elements~\cite{b5}. The transition metal element is incorporated in the alkali chalcogenide compound in the form of a solid solution. The substitution of the alkali metal with the $d$ or $f$ transition element is performed at up to about 25\% through a non-equilibrium crystal growth process at a low temperature to provide a ferromagnetic characteristic thereto.

Taking into account the importance of these materials due to their practical application, we reach the conclusion that a more precise calculation of the parameters of the electron energy spectra for them is an actual problem. And now let us turn to the solution.

\section{Calculation}


The first stage is to calculate the electron energy spectrum and eigenfunction in the generalized gradient approximation (GGA). For this purpose, the Kohn-Sham equations (\ref{1}) are solved in a self-consistent way~\cite{b7,b8}:
\begin{equation}\label{1}
\left(-\nabla^2+V_{\mathrm{ext}}+V_{\mathrm{H}}+V_{\mathrm{xc}}\right)\psi_{n\mathbf{k}}^{\mathrm{GGA}}(\mathbf{r})=\varepsilon_{n\mathbf{k}}^{\mathrm{GGA}}\psi_{n\mathbf{k}}^{\mathrm{GGA}}(\mathbf{r}),
\end{equation}
where $-\nabla^2$ is the kinetic energy operator, $V_{\mathrm{ext}}$ denotes the ionic pseudopotential, $V_{\mathrm{H}}$ and $V_{\mathrm{xc}}$ are the Hartree and exchange-correlation potential, respectively. The quasiparticle energies $\varepsilon_{n\mathbf{k}}^{qp}$  and eigenfunctions $\psi_{n\mathbf{k}}^{qp}(\mathbf{r})$ can be obtained from the quasiparticle equation~\cite{b9,b10}:
\begin{equation}\label{2}
\left(-\nabla^2+V_{\mathrm{ext}}(\mathbf{r})+V_{\mathrm{H}}(\mathbf{r})\right)\psi_{n\mathbf{k}}^{qp}(\mathbf{r})+ \int\Sigma\left(\mathbf{r},\mathbf{r'},\varepsilon_{n\mathbf{k}}^{qp}\right)\psi_{n\mathbf{k}}^{qp}(\mathbf{r'})\rd\mathbf{r'}= \varepsilon_{n\mathbf{k}}^{qp}\psi_{n\mathbf{k}}^{qp}(\mathbf{r}),
\end{equation}
where $\Sigma(\mathbf{r},\mathbf{r'},\varepsilon_{n\mathbf{k}}^{qp})$ is the non-local self-energy operator. The wave functions can be expanded as follows:
\begin{equation}\label{3}
|\psi_{n\mathbf{k}}^{qp}\rangle=\sum_{n'}\emph{a}_{n'}^n|\psi_{n\mathbf{k}}^{\mathrm{GGA}}\rangle.
\end{equation}
From equations~(\ref{2}) and (\ref{3}) the perturbative quasiparticle Hamiltonian is obtained in the form
\begin{equation}\label{4}
H_{nn'}^{qp}(E)=\varepsilon_{n\mathbf{k}}^{\mathrm{GGA}}\delta_{nn'}+\left\langle{ \psi_{n\mathbf{k}}^{\mathrm{GGA}}}|\Sigma(E)-V_{\mathrm{xc}}|\psi_{n'\mathbf{k}}^{\mathrm{GGA}}\right\rangle,
\end{equation}
where the perturbation is written as $\Sigma(E)-V_{\mathrm{xc}}$.

We have generated the PAW functions for the following valence basis states: ${1s^22s^12p^0}$ for Li, ${2s^22p^63s^13p^0}$ for Na, ${3s^23p^64s^14p^0}$ for K, ${4s^24p^65s^15p^0}$ for Rb, ${2s^22p^4}$ for O, ${2s^22p^63s^23p^4}$ for S, ${3s^23p^64s^24p^4}$ for Se, and ${4s^24p^65s^25p^4}$ for Te. All the PAW basis functions were obtained using the program \emph{atompaw}~\cite{b11}. The radii of the augmentation spheres are 1.6, 1.65, 2.3, 2.6, 1.45, 1.4, 1.8, and 2.4~a.u. for Li, N, K, Rb, O, S, Se, and Te, respectively. The values of the experimental lattice constants of the crystals A$_2$B used in calculations equal~\cite{b1} 8.642, 10.488, 12.170, and 12.741~a.u. for Li$_2$O, Na$_2$O, K$_2$O, and Rb$_2$O, respectively; 10.790, 12.332, 13.967, and 14.456~a.u. for Li$_2$S, Na$_2$S, K$_2$S, and Rb$_2$S, respectively; 11.342, 12.894, 14.967, and 15.154~a.u. for Li$_2$Se, Na$_2$Se, K$_2$Se, and Rb$_2$Se, respectively; 12.315, 13.850, 15.435, and 16.044~a.u. for Li$_2$Te, Na$_2$Te, K$_2$Te, and Rb$_2$Te, respectively.

The electronic energy bands and DOS have been evaluated by means of the ABINIT code~\cite{b12}. Integration over the Brillouin zone was performed on the Monkhorst-Pack~\cite{b13} grid of $6\times6\times6$ and $8\times8\times8$ in the GWA and GGA calculation, respectively. The iterations were performed to ensure the calculation of the total energy of the crystal with an accuracy of $10^{-8}$~Ha. The symmetry of the considered M$_2$A crystals is described by the space group Fm$\bar{3}$m (number 225), and the Bravais lattice is cF (face-centered cubic).

\section{Electronic properties}

\begin{figure}[!ht]
\includegraphics[width=0.48\textwidth]{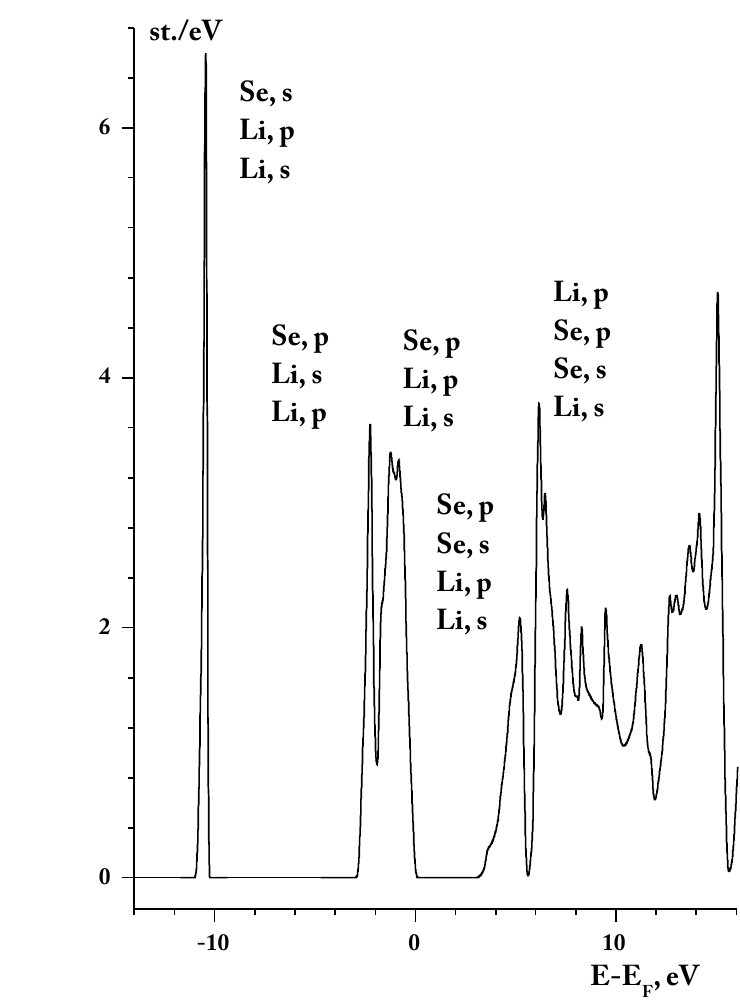}%
\hfill%
\includegraphics[width=0.48\textwidth]{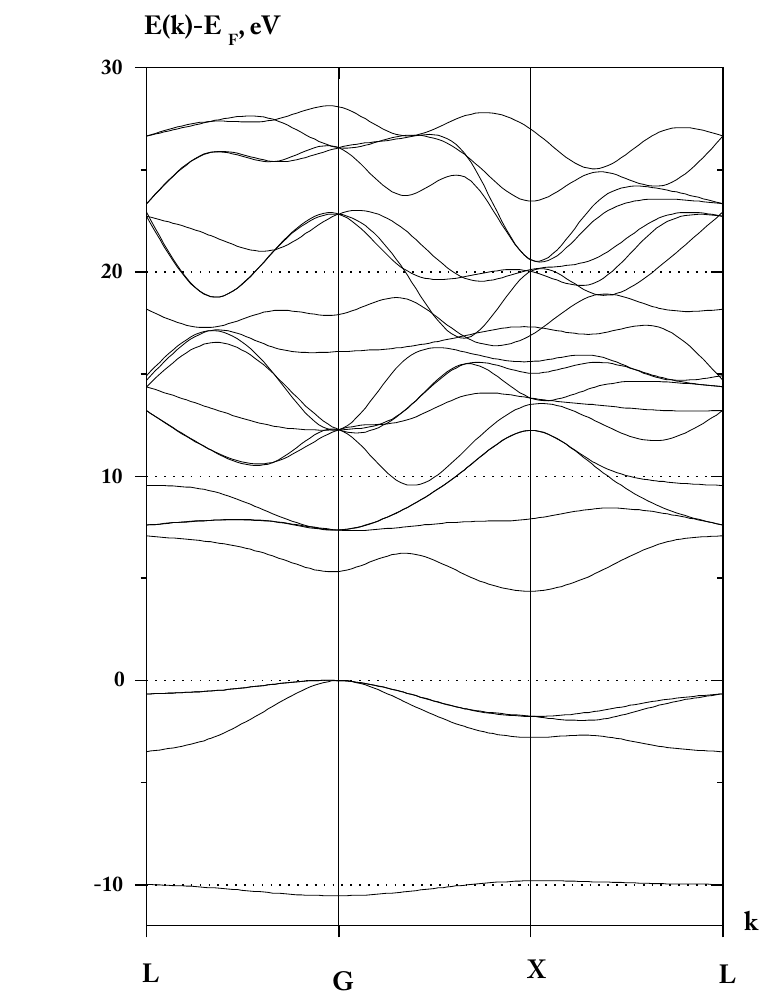}%
\\%
\parbox[t]{0.48\textwidth}{%
\caption{The total DOS of Li$_2$Se obtained in the GGA.}\label{f1}%
}%
\hfill%
\parbox[t]{0.48\textwidth}{%
\caption{The band structure of Li$_2$Se obtained in the GWA.}\label{f2}%
}%
\\[2ex]
\includegraphics[width=0.48\textwidth]{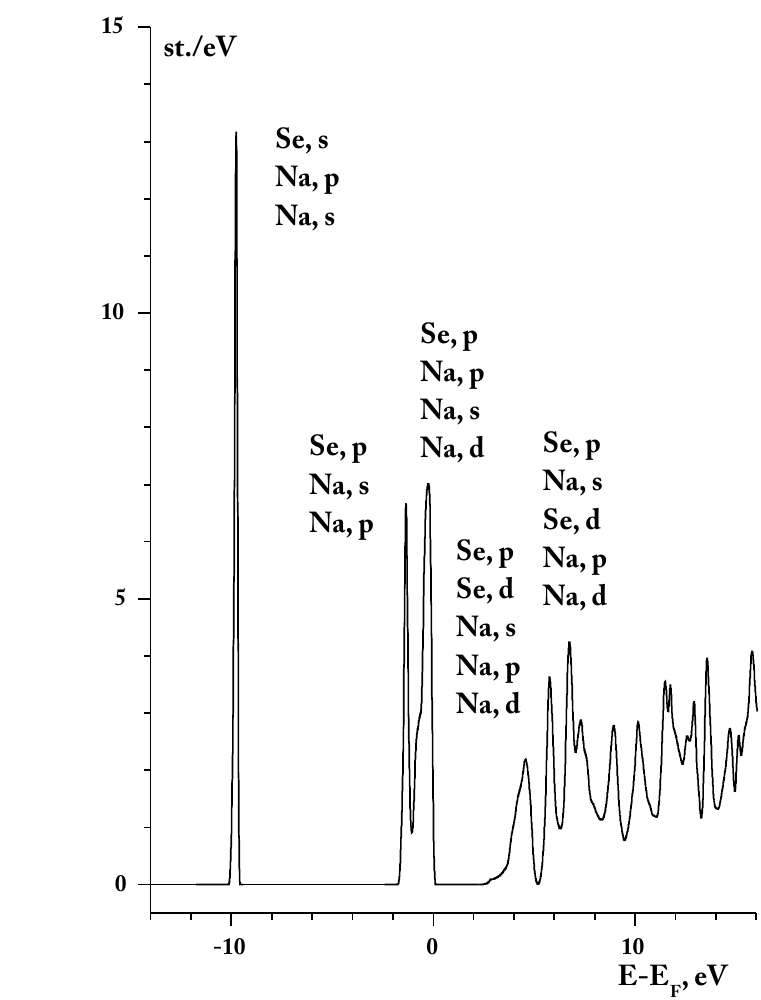}%
\hfill%
\includegraphics[width=0.48\textwidth]{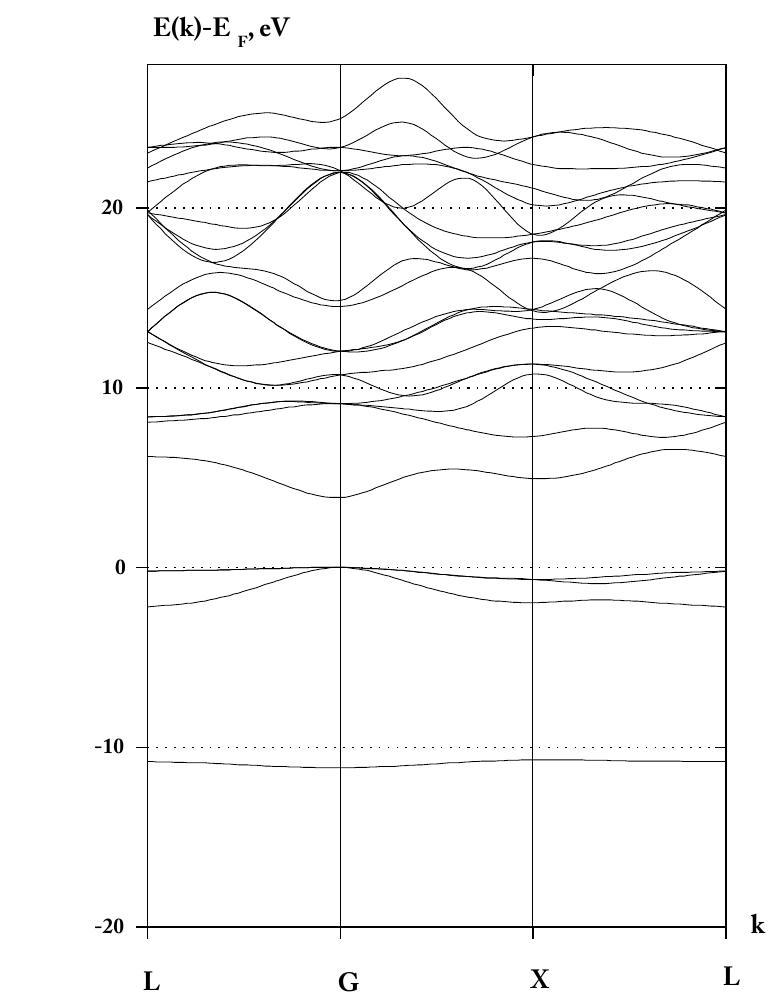}%
\\%
\parbox[t]{0.48\textwidth}{%
\caption{The total DOS of Na$_2$Se obtained in the GGA.}\label{f3}%
}%
\hfill%
\parbox[t]{0.48\textwidth}{%
\caption{The band structure of Na$_2$Se obtained in the GWA.}\label{f4}%
}%
\vspace{-5mm}
\end{figure}

Total density of electronic states (DOS) of  Li$_2$Se crystal is shown in figure~\ref{f1}. As can be seen, the wave functions of electrons in all energy bands are hybridized. This is indicated by the mark appearing next to the peaks on the DOS curves. For example, in the bottom of the valence band of the  Li$_2$Se  crystal, the $s$-states of Se dominate and the contributions to the DOS of $p$ and $s$ electrons of Li are less significant.
The dispersion curves, shown in figure~\ref{f2}, indicate that the crystal Li$_2$Se is a semiconductor with an indirect gap $\Gamma$-X. Figure~\ref{f1} shows the electronic DOS of the  Li$_2$Se crystal, evaluated within the GGA on the PAW basis. As can be seen, the bottom of the valence band is characterized by a small dispersion, and the corresponding curve is localized in a narrow strip about 0.53~eV wide. The value of the corresponding parameter obtained within GWA (see figure~\ref{f2}) is slightly greater and equals 0.72~eV. The widths of the upper parts of the valence bands are characterized by the values of 2.91 and 3.49~eV obtained in the GGA and GWA, respectively. The width of the valence band found in the GGA equal 10.90~eV, and the corresponding value obtained in the GWA is 10.53~eV.

Now, let us analyze the results of the calculation obtained for the  Na$_2$Se crystal. Let us consider the DOS in figure~\ref{f3} and the dispersion curves in figure~\ref{f4}. They show that this crystal is characterized by a direct gap at the point $\Gamma$. As can be seen from figure~\ref{f3} (GGA), the bottom of the valence band is characterized by a small dispersion and the corresponding curve is localized in a narrow strip about 0.29~eV wide. Analogous parameter obtained in the GWA (figure~\ref{f4}) is a little greater and equals 0.44~eV. The values of the widths of the upper part of the valence band obtained within the GGA and GWA, are equal to 1.71 and 2.18~eV, respectively. The total width of the valence band found in the GGA equals 10.04~eV, and the corresponding value obtained in the GWA is 11.13~eV.

\begin{figure}[!ht]
\vspace{-5mm}
\includegraphics[width=0.48\textwidth]{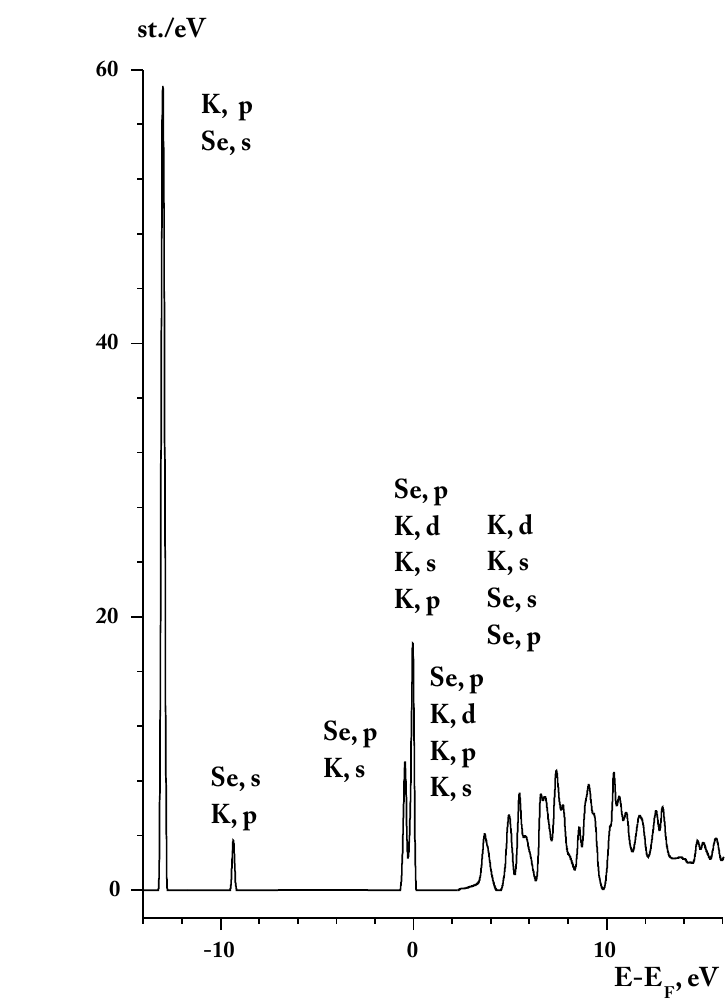}%
\hfill%
\includegraphics[width=0.48\textwidth]{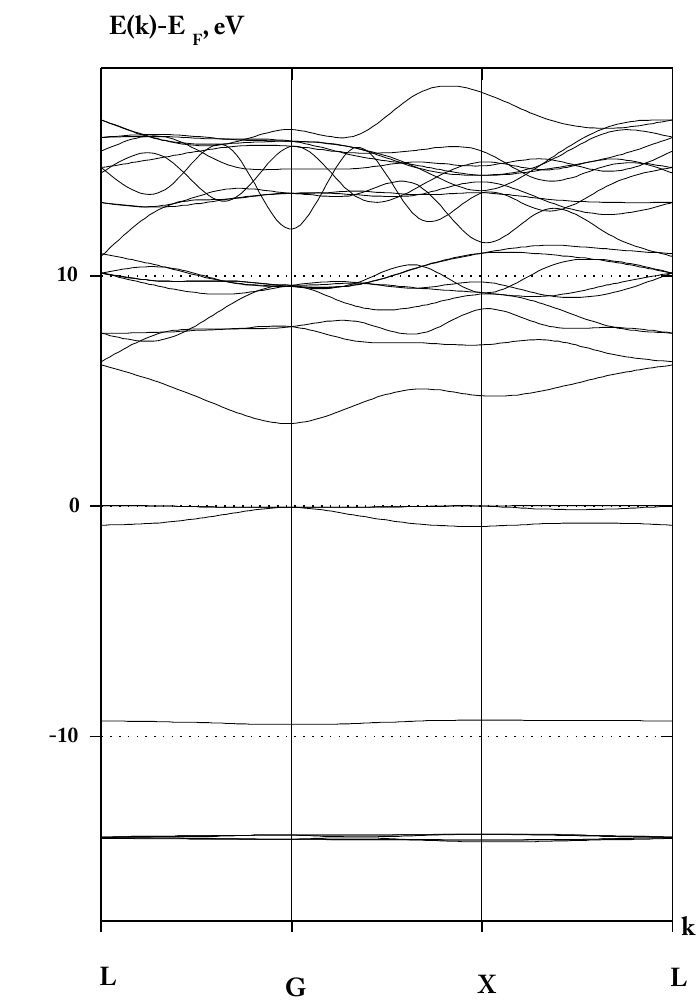}%
\\%
\parbox[t]{0.48\textwidth}{%
\caption{The total DOS of K$_2$Se obtained in the GGA.}\label{f5}%
}%
\hfill%
\parbox[t]{0.48\textwidth}{%
\caption{The band structure of K$_2$Se obtained in the GWA.}\label{f6}%
}%
\\[2ex]
\includegraphics[width=0.48\textwidth]{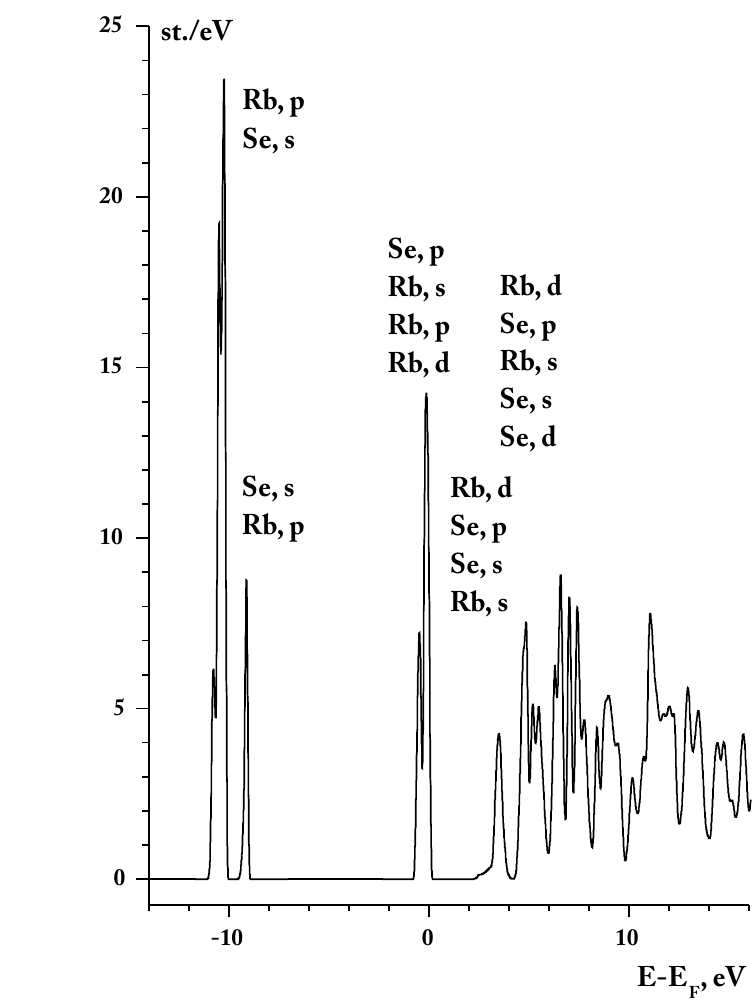}%
\hfill%
\includegraphics[width=0.48\textwidth]{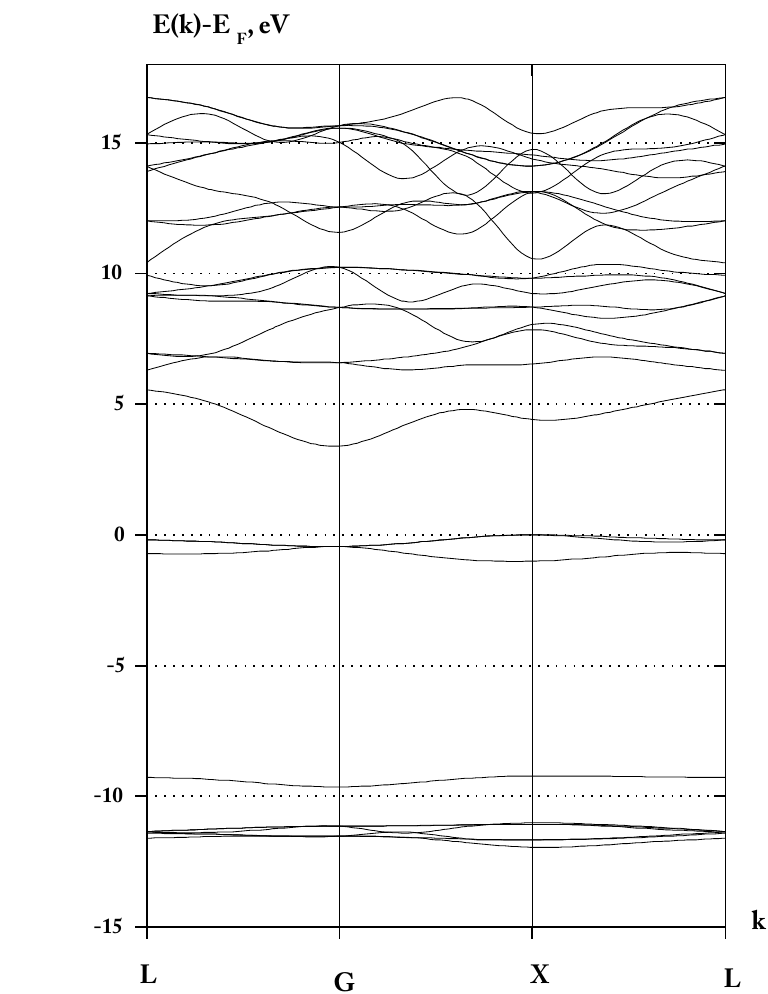}%
\\%
\parbox[t]{0.48\textwidth}{%
\caption{The total DOS of Rb$_2$Se obtained in the GGA.}\label{f7}%
}%
\hfill%
\parbox[t]{0.48\textwidth}{%
\caption{The band structure of Rb$_2$Se obtained in the GWA.}\label{f8}%
}%
\vspace{-5mm}
\end{figure}

Figures~\ref{f5} and \ref{f6} show the electronic energy bands of the  K$_2$Se crystal evaluated within the GGA and GWA, respectively. As can be seen, the crystal has an indirect gap $\Gamma$-X. The lowest bands calculated within the GGA and GWA are localized in very narrow strips of 0.27 and 0.34~eV, respectively. They consist of the core states of the K atom. The bottom of the valence band corresponding to the GGA and GWA is localized within a very narrow strip of about 0.14 and 0.19~eV wide, respectively. The strips containing the upper parts of the valence band calculated by means of the GGA and GWA are 0.66 and 0.90~eV wide, respectively. The total width of the valence band obtained in the GGA and GWA is 9.41 and 9.49~eV, respectively.

At last, let us turn to the analysis of the results found for the crystal Rb$_2$Se represented in figures~\ref{f7} and \ref{f8}. As can be seen from figures~\ref{f7} and \ref{f8},  the  Rb$_2$Se crystal has an indirect gap $\Gamma$-X. In the lowest bands,  the core p-states of Rb dominate. They lie in a strip of the width of 0.80 and 0.92~eV, respectively. The bottom of the valence band is created by $s$-states of selenium. It is located just above the core $p$-states of rubidium. The widths of the bottoms of the valence bands are 0.407 and 0.413~eV obtained within the GGA and GWA, respectively. The top of the valence band consists mainly of $s$- and $p$-states of selenium. The widths of the corresponding bands are equal to 0.75 and 1.00~eV, respectively. Finally, the full width of the valence band, obtained in the GGA and GWA, equals 9.52 and 9.64~eV, respectively.

Now, consider the results of the calculation presented in table~\ref{tbl}. The values of the band energies obtained in the FP APW~\cite{b1} approach by means of the WIEN2K code, and evaluated here in the PBE PAW framework with ABINIT code, are substantially underestimated. Let us first consider the properties of the Li$_2$O crystal  for which the experimental value of the energy of the optical absorption is known~\cite{b4}. The value of the X${}-\Gamma$ gap, found in~\cite{b1} within the DFT is 4.96~eV, and our value equals 5.07~eV. And the value of this parameter, calculated here within the GWA equals 7.55~eV. However, the experimental value of the optical absorption energy is equal to 6.6~eV. Now it is possible to estimate the binding energy of an exciton, which is simply equal to the difference of the last two values of the energy that is 0.95~eV. The corresponding value found recently from the Bethe-Salpeter equation is 0.98~eV~\cite{b14}.

Table~\ref{tbl} shows that all the  Li$_2$A, K$_2$A and Rb$_2$A crystals have an indirect band gap X${}-\Gamma$, and Na$_2$A  crystals are characterized by a direct gap $\Gamma-\Gamma$. Table~\ref{tbl} shows that the values of the direct and indirect gaps in the  Li$_2$A  crystals monotonously decrease with the replacement of the second element $\rm O\rightarrow S \rightarrow Se \rightarrow Te$. A similar behavior is also shown by the direct gap X${}-{}$X in the~crystals~Na$_2$A.

Now, consider the results of the calculation presented in table~\ref{tb2}. As can be seen, the most significant changes in the energy gaps $\Delta E$ are obtained for the crystal Li$_2$O. We pay attention to the fact that the values of all the changes in the energy gaps $\Delta E$ obtained for each crystal are different. The greatest value of the $\Delta E$ change for the direct gap $\Gamma-\Gamma$ is obtained for the Li$_2$O crystal and the smallest value is found for the  Li$_2$Te crystal.

The macroscopic dielectric function $\varepsilon^{\mathrm{LF}}_{\mathrm{M}}(\omega)$, including local field effects, is related to the inverse of the microscopic dielectric matrix ~\cite{b12}: $\varepsilon^{\mathrm{LF}}_{\mathrm{M}}(\omega)=\lim\limits_{\mathbf{q}\rightarrow0}\frac{1}{\varepsilon_{00}^{-1}\left(\mathbf{q},\omega\right)}$. If local fields are neglected (no local fields, NLF), the irreducible polarizability is computed in the independent particle approximation. In this case,
$\varepsilon^{\mathrm{NLF}}_{\mathrm{M}}(\omega)=\lim\limits_{\mathbf{q}\rightarrow0}{\varepsilon_{00}\left(\mathbf{q},\omega\right)}$. The value $\varepsilon_{\mathrm{M}}(0)$  is the static dielectric constant $\varepsilon_{\infty}$  presented in table~\ref{tb3}. The value of the dielectric constant for the Li$_2$O crystal obtained here is 2.65, and the one evaluated in work ~\cite{b16} is 2.62. The corresponding experimental result equals 2.68 ~\cite{b4}. The convergence of the values of dielectric constants listed in table~\ref{tb3} served as an additional criterion of the choice of the plane wave basis in the Kohn-Sham problem, and in the calculation of the exchange $\Sigma_\emph{x}$  and correlation $\Sigma_\emph{c}$  parts ~\cite{b12} of the self-energy.

\begin{table}[htb]
\caption{The calculated electronic band gaps of the crystals M$_2$A, in~eV.}
\label{tbl}
\begin{center}
\begin{tabular}{|*{13}{c|}}\hline
\multicolumn{1}{|c|}{} & \multicolumn{4}{c|}{\rm PBE~FP~APW~\cite{b1}} & \multicolumn{4}{c|}{$E_{\mathrm{GGA}}$} & \multicolumn{4}{c|}{$E_{\mathrm{GWA}}$} \\ \hline
$\rm Li_2A$ & O & S & Se & Te & O & S & Se & Te & O & S & Se & Te \\ \hline\hline
$\Gamma-\Gamma$ & 5.15 & 4.19 & 3.45 & 3.19 & 5.52 & 4.26 & 3.67 & 3.79 & 8.46 & 6.03 & 5.32 & 4.05 \\ \hline
$\rm X-\Gamma$ & 4.96 & 3.36 & 2.93 & 2.46 & 5.07 & 3.47 & 3.04 & 2.59 & 7.55 & 4.73 & 4.36 & 3.59 \\ \hline
$\rm X-X$ & 6.31 & 4.77 & 4.36 & 3.69 & 6.48 & 4.88 & 4.48 & 4.05 & 9.35 & 6.54 & 6.12 & 5.35 \\ \hline
$\rm Na_2A$ &  &  &  &  &  &  &  &  &  &  &  & \\ \hline
$\Gamma-\Gamma$ & 1.83 & 2.40 & 2.09 & 2.11 & 2.00 & 2.56 & 2.25 & 2.51 & 3.93 & 4.24 & 3.90 & 4.00 \\ \hline
$\rm X-\Gamma$ & 4.61 & 2.85 & 2.58 & 2.72 & 4.74 & 3.93 & 3.55 & 3.13 & 6.48 & 5.24 & 4.94 & 4.28 \\ \hline
$\rm X-X$ & 4.86 & 4.27 & 3.96 & 3.93 & 4.98 & 4.37 & 4.06 & 3.72 & 6.81 & 5.82 & 5.59 & 5.02 \\ \hline
$\rm K_2A$ &  &  &  &  &  &  &  &  &  &  &  & \\ \hline
$\Gamma-\Gamma$ & 5.14 & 2.40 & 2.32 & 2.28 & 2.34 & 2.68 & 2.19 & 2.60 & 3.73 & 4.10 & 3.65 & 4.02 \\ \hline
$\rm X-\Gamma$ & 1.71 & 2.24 & 2.03 & 2.02 & 1.86 & 2.47 & 2.11 & 2.57 & 3.09 & 3.82 & 3.58 & 4.00 \\ \hline
$\rm X-X$ & 3.23 & 3.41 & 3.22 & 2.94 & 3.22 & 3.54 & 3.36 & 3.22 & 4.59 & 4.83 & 4.78 & 4.54 \\ \hline
$\rm Rb_2A$ &  &  &  &  &  &  &  &  &  &  &  & \\ \hline
$\Gamma-\Gamma$ & 1.88 & 2.28 & 2.21 & 2.18 & 2.40 & 2.73 & 2.42 & 2.56 & 3.59 & 4.03 & 3.85 & 3.93 \\ \hline
$\rm X-\Gamma$ & 1.31 & 1.94 & 1.88 & 1.96 & 1.78 & 2.37 & 2.08 & 2.33 & 2.62 & 3.50 & 3.40 & 3.62 \\ \hline
$\rm X-X$ & 2.69 & 3.11 & 3.15 & 3.02 & 2.97 & 3.40 & 3.14 & 3.04 & 3.59 & 4.45 & 4.42 & 4.24 \\ \hline
\end{tabular}
\renewcommand{\arraystretch}{1}
\end{center}
\end{table}
\begin{table}[!h]
\caption{The differences between the energy gaps of the crystals M$_2$A evaluated within the GWA and GGA approaches.}
\label{tb2}
\begin{center}
\begin{tabular}{|*{9}{c|}}\hline
\multicolumn{1}{|c|}{} & \multicolumn{4}{c|}{$\Delta E=E_{\mathrm{GWA}}-E_{\mathrm{GGA}}, \rm eV$} & \multicolumn{4}{c|}{$\Delta E/E_{\mathrm{GWA}}$} \\ \hline\hline
$\rm Li_2A$ & O & S & Se & Te & O & S & Se & Te \\ \hline
$\Gamma-\Gamma$ & 2.94 & 1.77 & 1.65 & 0.26 & 0.35 & 0.29 & 0.31 & 0.06 \\ \hline
$\rm X-\Gamma$ & 2.48 & 1.26 & 1.32 & 1.00 & 0.33 & 0.27 & 0.30 & 0.28 \\ \hline
$\rm X-X$ & 2.87 & 1.66 & 1.64 & 1.30 & 0.31 & 0.25 & 0.27 & 0.24 \\ \hline
$\rm Na_2A$ &  &  &  &  &  &  &  &  \\ \hline
$\Gamma-\Gamma$ & 1.93 & 1.68 & 1.65 & 1.49 & 0.49 & 0.40 & 0.42 & 0.37 \\ \hline
$\rm X-\Gamma$ & 1.74 & 1.31 & 1.39 & 1.15 & 0.27 & 0.25 & 0.28 & 0.27 \\ \hline
$\rm X-X$ & 1.83 & 1.45 & 1.53 & 1.30 & 0.27 & 0.25 & 0.27 & 0.26 \\ \hline
$\rm K_2A$ &  &  &  &  &  &  &  &  \\ \hline
$\Gamma-\Gamma$ & 1.39 & 1.42 & 1.46 & 1.42 & 0.37 & 0.35 & 0.40 & 0.35 \\ \hline
$\rm X-\Gamma$ & 1.23 & 1.35 & 1.47 & 1.43 & 0.40 & 0.35 & 0.41 & 0.36 \\ \hline
$\rm X-X$ & 1.37 & 1.29 & 1.42 & 1.32 & 0.30 & 0.27 & 0.30 & 0.29 \\ \hline
$\rm Rb_2A$ &  &  &  &  &  &  &  & \\ \hline
$\Gamma-\Gamma$ & 1.19 & 1.30 & 1.43 & 1.37 & 0.33 & 0.32 & 0.37 & 0.35 \\ \hline
$\rm X-\Gamma$ & 0.84 & 1.13 & 1.32 & 1.29 & 0.32 & 0.32 & 0.39 & 0.36 \\ \hline
$\rm X-X$ & 0.62 & 1.05 & 1.28 & 1.20 & 0.17 & 0.24 & 0.29 & 0.28 \\ \hline
\end{tabular}
\renewcommand{\arraystretch}{1}
\end{center}
\end{table}

\begin{table}[htb]
\caption{The calculated dielectric constants for the crystals M$_2$A found with and without the local field (LF) effects.}
\label{tb3}
\begin{center}
\begin{tabular}{|*{9}{c|}}\hline
\multicolumn{1}{|c|}{} & \multicolumn{4}{c|}{$\varepsilon_{\infty}, \rm with~LF$} & \multicolumn{4}{c|}{$\varepsilon_{\infty}, \rm without~LF$} \\ \hline
$\rm A$ & O & S & Se & Te & O & S & Se & Te \\ \hline\hline
$\rm Li_2A$ & 2.65 & 3.65 & 3.88 & 4.17 & 2.89 & 4.47 & 4.72 & 5.09 \\ \hline
$\rm Na_2A$ & 2.98 & 3.24 & 3.43 & 3.54 & 3.26 & 3.96 & 4.19 & 4.38 \\ \hline
$\rm K_2A$ & 2.98 & 2.94 & 2.93 & 2.96 & 3.49 & 3.72 & 3.72 & 3.81 \\ \hline
$\rm Rb_2A$ & 3.91 & 2.86 & 2.94 & 2.89 & 4.57 & 3.59 & 3.73 & 3.72 \\ \hline
\end{tabular}
\renewcommand{\arraystretch}{1}
\end{center}
\end{table}

\section{Conclusions}

The electron energy spectra for M$_2$A crystals have been originally calculated based on quasiparticle corrections within the GW approach. The results obtained herein show that the values of the interband gaps found without the quasiparticle corrections are usually underestimated by $20-50$ percent (see table~\ref{tb2}). All the Na$_2$A crystals considered here are characterized by direct gaps $\Gamma-\Gamma$. The rest of the M$_2$A crystals have indirect gaps X${}-\Gamma$. The non-local self-energy operator $\Sigma$ in equation~(\ref{2}) was evaluated without application of the plasmon pole model. The GW calculations have been carried out using the ABINIT code employing the contour deformation method~\cite{b12,b15}. As can be seen from table~\ref{tb2},  the corrections $\Delta E$ are not weakly dependent on the wave vector. Therefore, the scissor operator is not a good approximation for all the crystals considered here.  The long wave limits of the dielectric constants of the considered crystals have been evaluated for the first time. The last one found for  Li$_2$O crystal is well compared with the experimental value. Table~\ref{tb3} shows that the nine crystals listed therein have dielectric constants less than~3.0. We can assume that the exciton binding energy possessed by them is in the range from about~0.5 to 1.0~eV. Thus, the bandgap calculated in the GWA would exceed the experimental value of the optical absorption energy by the value of the binding energy of the exciton~\cite{b4,b14}. We hope that the results obtained here will stimulate the experimental study of these materials, which is important for practical applications.

%

\ukrainianpart

\title{Квазічастинкова електронна енергетична структура халькогенідів лужних металів}
\author{С.В. Сиротюк, В.М. Швед}
\address{Національний університет ``Львівська політехніка'', вул. С.~Бандери, 12, 79013 Львів, Україна}
%
%
%
\makeukrtitle

\begin{abstract}
\tolerance=3000%
Електронні енергетичні спектри халькогенідів лужних металів M$_2$A (M: Li, Na, K, Rb; O, S, Se, Te)
були підраховані за методом проекційних приєднаних хвиль (PAW) за допомогою програми ABINIT.
Одночастинкові стани у формалізмі Кона-Шема були знайдені в рамках GGA (узагальнене градієнтне наближення).
Далі на основі цих результатів були отримані квазічастинкові енергії електронів
та діелектричні константи у наближенні GW. Для розглянутих кристалів M$_2$A розрахунки
на основі функції Гріна були зроблені вперше, за винятком Li$_2$O.
\keywords електронна структури, формалізм GGA, формалізм GWA, метод проекційних приєднаних хвиль, діелектрична стала

\end{abstract}

\end{document}